\documentclass[prl,amsmath,amssymb,nofootinbib,showpacs,twocolumn]{revtex4-1}
\usepackage{textcomp}

\usepackage{graphicx}
\usepackage{amsmath,amssymb,epsfig}
\newcommand{\cc}{\mbox{c.c.}}

\newcommand{\dd}{\partial}

\newcommand{\pd}[2]{\frac{\partial #1}{\partial #2}}

\newcommand{\eps}{\epsilon}

\newcommand{\LL}{\mathcal{L}}
\newcommand{\NN}{\mathcal{N}}

\newcommand{\beqa}{\begin{eqnarray}}
\newcommand{\eeqa}{\end{eqnarray}}
\newcommand{\beqas}{\begin{eqnarray*}}
\newcommand{\eeqas}{\end{eqnarray*}}
\newcommand{\beq}{\begin{equation}}
\newcommand{\eeq}{\end{equation}}

\newcommand{\bal}{\begin{align}}
\newcommand{\ealign}{\end{align}}

\newcommand{\ii}{\mathrm{i}}
\newcommand{\ve}[1]{\mathbf{#1}}

\begin{document}

\title{Analytical results for front pinning between an hexagonal pattern and a uniform state in pattern-formation systems}
\author{G. Kozyreff$^1$ and S.J. Chapman$^2$}
\affiliation{$^1$Optique Nonlin\'eaire Th\'eorique, Universit\'e libre de Bruxelles (U.L.B.), CP 231, Belgium}
\affiliation{$^2$OCIAM, Mathematical Institute, 24-29 St Giles', Oxford OX13LB, UK}
\begin{abstract}
In pattern-forming systems, localized patterns are states of intermediate complexity between fully extended ordered patterns and completely irregular patterns. They are formed by stationary fronts enclosing an ordered pattern inside an homogeneous background. In two dimensions, the ordered pattern is most often hexagonal and the conditions for fronts to stabilize are still unknown. In this letter, we show how the locking of these fronts depends on their orientation relative to the pattern. The theory rests on general asymptotic arguments valid when the spatial scale of the front is slow compared to that of the hexagonal pattern. Our analytical results are confirmed by numerical simulations with the Swift-Hohenberg equation, relevant to hydrodynamical and buckling instabilities, and a nonlinear optical cavity model.
\end{abstract}
\date{\today}
\maketitle
Pattern-forming instabilities are responsible for the emergence of regular, or semi-regular structures in a wide variety of contexts. Examples include: chemical Turing patterns \cite{Ouyang-1991}, hydrodynamic convection \cite{Schatz-1995,Bergeon-2008}, ferro-solitons \cite{Richter-2005}, cavity solitons \cite{Barbay-2008}, and buckling instabilities \cite{Vandeparre-2008,Wadee-2012}. Usually, patterns bifurcate from an homogeneous state and lead to periodic spatial modulation of the quantity of interest (chemical concentrations, velocity fields, optical intensity). If both the pattern and the homogeneous state coexist, then there exists a range of parameters for which islands of patterns are embedded in an homogeneous background. Recently, localized convection patterns were discovered in planar Couette flow and a connection with localized turbulence was speculated \cite{Schneider-2010}. Such a situation, while very rich from a dynamical point of view, is still amenable to a simple description in terms of fronts connecting the pattern and the homogeneous state. It is therefore important to understand under which conditions such fronts can be stationary. A lot of progress has already been made towards the understanding of these fronts in one dimension (1D). They are based, on the one hand, on geometrical arguments in phase space \cite{Hunt-1999,Coullet-2000,Beck-2009} and, on the other hand, on beyond-all-order asymptotic techniques \cite{Kozyreff-2006,Chapman-2009,Dean-2011}. The latter allows one to properly describe front locking -- also called `front pinning' \cite{Pomeau-1984,Pomeau-1986}--  through the interaction between the slow spatial scale of the front and the fast spatial scale of the pattern. Thanks to these and other approaches involving asymptotic and numerical studies \cite{Bensimon-1988,Burke-2006a,Burke-2007b,Wadee-1999,Wadee-2002,Susanto-2011}, a coherent picture is now emerging for  1D localized patterns.

For  localized patterns on the plane (2D), research is intensifying. The pinning region  of localized hexagonal patterns was computed numerically in \cite{McSloy-2002,Lloyd-2008}; a detailed analysis of radial patterns was conducted in \cite{Lloyd-2009}; meanwhile 2D localized roll patterns and `worm' patterns were studied in \cite{Hagberg-2006,Burke-2007b,Avitabile-2010}. Regarding localized hexagonal patterns however, the amount of available analytical information past the mere writing of the Ginzburg-Landau equations for the envelope of the pattern is still scarce. Indeed these equations are too complicated even to find analytical front solutions in general \cite{Malomed-1990,Pismen-2006}. Nevertheless, the method used in \cite{Kozyreff-2006,Chapman-2009,Dean-2011} can be carried sufficiently far in 2D to relate the size of the pinning region to the front orientation for a broad class of pattern-formation models in the small-amplitude limit. We will show that fronts which are parallel to one side of the elementary hexagon of the pattern have the widest range of existence. The next widest range of existence is found for fronts that are perpendicular to one side of the elementary hexagon. In fact, a complete hierarchy of front pinning ranges versus orientation can be established. Since in many cases fronts make up the boundary of localized patterns, this sets strong geometrical constraints on the shape of 2D localized patterns. These findings will be illustrated with two physical models:
\beq
\pd ut=ru-\left(1+\nabla^2\right)^2u+su^2-u^3,
\label{SH}
\eeq
and 
\beq
\pd Et=E_I-\left(1+\frac{2C}{1+|E|^2} \right)E+\ii\left(\theta+\nabla^2\right)E.
\label{NLO}
\eeq
Equation (\ref{SH}) is the prototype model for pattern formation. Initially derived (with a cubic nonlinearity only) for convective instabilities \cite{Swift-1977}, it is the simplest model for the Turing instability and is widely used to interpret more complicated and physically realistic problems \cite{Cross-1993,Knobloch-open}. Equation (\ref{NLO}) describes the envelope of the electric field in the transverse plane of a nonlinear  optical cavity \cite{McSloy-2002}. Above, $E_I$ is an injection field amplitude, $\theta$ is the cavity detuning and $C$ is the strength of the light-matter interaction. If one expresses $E_I$ as
\beq
E_I=\left| 1+\ii\theta+\frac{2C}{1+I}  \right|I^{1/2},
\eeq
then the homogeneous steady state is simply given by $|E|^2=I$. For the sake of simplicity, we will fix the value $\theta=1$ and use $C$ as a parameter.  Both models above exhibit a Turing instability and can display localized patterns in proper regions of parameter space: $(r,s)$ in Eq.~(\ref{SH}) and $(I,C)$ in Eq.~(\ref{NLO}). The aim of this paper is to determine the size of these existence regions as a function of the orientation of the pattern boundaries.

Upon elementary changes of variable, Eqs.~(\ref{SH}) and (\ref{NLO}) can be put, in steady state, in the general form
\beq
\LL(\ve u, \nabla^2 \ve u, \nabla^4\ve u,\ldots;\eps,\nu)+ \NN(\ve u;\eps,\nu)=\ve 0,
\label{general}
\eeq
where $\ve u(x,y)$ is a vector field (set of chemical concentrations, electric field,
 elastic displacement field), $\LL$ is a linear operator involving  $\nabla^2, \nabla^4$, or higher-order composition thereof, and $\NN$
 is the nonlinear part of the mathematical model, being at least 
quadratic in $\ve u$ as $\ve u\to \ve 0$. Finally, $(\eps, \nu)$ are two control parameters. Without loss of generality, we may assume
 that the spatially uniform solution of (\ref{general}) is given by
 $\ve u=\ve 0$ and that $\eps=0$ is the bifurcation point to spatially
 periodic solutions. Hence, we wish to derive an expression for the width
 of the pinning region, $\delta \nu(\eps)$, when $0<\eps\ll1$. 

We begin by recalling general results \cite{Cross-1993,Pismen-2006}. At the bifurcation point, there is an eigenvector $\ve u_T$ and a real wave number $k$ such that $\ve u_T \exp\left(\ii \ve k\cdot\ve x\right)$ solves the linear part of (\ref{general}):
\beq
\LL(\ve u_T, -k^2 \ve u_T, k^4\ve u_T, \ldots;0,\nu)=\ve 0,\qquad k^2=\ve k\cdot\ve k.
\label{linear}
\eeq
Any combination of plane waves of the form $\ve u_T\exp\left(\ii \ve k'\cdot\ve x\right)$ satisfies (\ref{linear}) as long as $\ve k'\cdot\ve k'=k^2$. What matters here is that
\beq
\nabla^2 e^{\ii \ve k\cdot\ve x}=-k^2 e^{\ii \ve k\cdot\ve x}.
\label{ksquare}
\eeq
Hexagonal patterns are composed of triads $\ve k_i, i=1,2,3$ of such wave vectors, which satisfy $\ve k_1+\ve k_2+\ve k_3=\ve0$. Following standard procedures, we seek an approximate solution of the form
\begin{align}
\ve u&\sim\eps \ve u_T\sum_{l=1}^3 a_l(X)e^{\ii\ve k_l\cdot\ve x}+\cc, &X&=\eps x, &\eps&\ll1,
\label{leading}
\end{align}
where $\cc$ means `complex conjugate' and where $x$ and $X$ are treated asymptotically as independent variables. To be able to describe a front, the amplitudes $a_i$ are allowed to vary slowly in the $x$-direction.  With proper definitions of $\eps$ and $\nu$, the amplitude equations have the general form\begin{multline}
4n_i^2a_i''=a_i\left(1+|a_i|^2+2|a_j|^2+2|a_k|^2\right)+\nu \bar a_j\bar a_k,
\label{GL}
\end{multline}
where $\{i,j,k\}$ is any permutation of $\{1,2,3\}$ and $n_i$ is the direction cosine between the vector $\ve k_i$ and the $x$-axis. In Eq.~(\ref{SH}), for instance, one should take  $r=-\eps^2$ and $s=\sqrt{3/4}\eps\nu$, while in Eq.~({\ref{NLO}), one must set $I=3+\sqrt{4/3}\eps \nu$ and $C=4+\eps^2(\nu^2-4)/3$.

These equations have the first integral
\begin{multline}
\sum_i\left[4n_i^2|a_i'|^2-|a_i|^2-\frac{|a_i|^4}2-\frac{2|a_1a_2a_3|^2}{|a_i|^{2}}\right]\\
-\nu\left(a_1a_2a_3+\bar a_1\bar a_2\bar a_3\right)
\label{integral}
\end{multline}
which implies that front solutions exist at $\nu=-\sqrt{45/2}$. The corresponding hexagon amplitude is $a_1=a_2=a_3=\sqrt{2/5}$. 
Additional terms proportional to  $n_j\bar a_j'\bar a_k+n_k\bar a_j\bar a_k'$ sometimes appear in (\ref{GL})
(see  \cite{Pismen-2006}.) They preclude the existence of a first integral. In this work, we focus on the frequent case where they are absent, so that the existence of a monotonic front solution can be ensured.

Taking $\nu=-\sqrt{45/2}$, fronts are thus expected from the above theory on the lines $(r,s)=(-\eps^2,\sqrt{135/8}\eps)$ and $(I,C)=(3-\sqrt{30}\eps,4+37\eps^2/6)$ in the two examples considered.  However, fronts are known to exist on a finite area of the parameter space --not just a line-- and we now proceed to determine it. 

In fact, expression (\ref{leading}) is only the first term of a more general asymptotic approximation
\beq
\ve u(x,y;\eps)=\sum_{n=1}^{N}\eps^n\ve u_n(x,y;X)+R_N(x,y;\eps).
\label{expansion}
\eeq
The power series above diverges as $N\to\infty$ and has therefore to be truncated. Truncating the series near its smallest term leaves the remainder $R_N$ exponentially small in $\eps$. It is the study of $R_N$ which gives access to the front dynamics \cite{Kozyreff-2006,Chapman-2009,Dean-2011}. 

$R_N$ is mainly determined by the terms in (\ref{expansion}) that diverge  most rapidly as $n\to\infty$. A typical cause of divergence of asymptotic expansions is the presence of complex singularities in the leading order solution \cite{Chapman-1998}. Denoting one such singularity by $X_0$, we expect that the $O(\eps^n)$ term of (\ref{expansion}) contains, among others, contributions of the form
\beq
\eps^nb_{\ve q}\frac{\Gamma(\alpha+n)e^{\ii\ve q\cdot\ve x}}{\left[-\ii\Delta k (X-X_0)\right]^n}\ve u_T,\qquad n\gg1,
\label{ansatz}
\eeq
where $\ve q=m_1\ve k_1+m_2\ve k_2$, $b_\ve q$, $\alpha$, and $\Delta
k$ are constants with $m_1$ and $m_2$ integers. The vector $\ve q$
arises from the mixing of $\ve k_1$, $\ve k_2$, and $\ve k_3$ due to
the nonlinearity, and can be written in terms of
$\ve k_1$ and $\ve k_2$ only since $\ve  k_3 = -\ve  k_1 - \ve k_2$. The above ansatz arises from condition (\ref{ksquare}) which determines the most unstable modes at the bifurcation point $\eps=0$. In the frame of the multiple-scale analysis, the laplacian operator is replaced by $\nabla^2\to\nabla^2+2\eps\dd_x\dd_X+\eps^2\dd_X^2$. Hence, at $O(\eps^n)$, $\nabla^2u$ yields $\nabla^2u_n+ 2\dd_x\dd_X u_{n-1}+\dd_X^2u_{n-2}$. Substituting (\ref{ansatz}), one easily sees that this yields
\begin{multline}
-\left[\left(q_x+\Delta k\right)^2+q_y^2+O(1/n)\right]
\frac{\eps^nb_{\ve q}\Gamma(\alpha+n)e^{\ii\ve q\cdot\ve x}\ve u_T}{\left[-\ii\Delta k (X-X_0)\right]^n}
\end{multline}
where $q_x$ and $q_y$ are the cartesian coordinates of $\ve q$ in the $(x,y)$ frame. With appropriate choices of $\ve q$ and $\Delta k$, one may have $\left(q_x+\Delta k\right)^2+q_y^2=k^2$. As a result, terms of the form (\ref{ansatz}) approximately solve the linear part of (\ref{general}) and we expect them to dominate the series (\ref{expansion}) for large $n$. Let the $x$ axis coincide with a direction of translational symmetry of the lattice spanned by $\ve k_1$ and $\ve k_2$; then $\Delta k$ is such that $(q_x+\Delta k,q_y)$ coincides with one of $\pm \ve k_i$. Note that, in that case, a countable infinity of allowed values exist for $\Delta k$, each in principle leading to a set of terms of the form (\ref{ansatz}). 

Terms like (\ref{ansatz}), although formally non resonant, may become resonant over a short portion of the real $X$ axis: This is where the interaction between the slow and fast spatial scales occurs, as discussed in detail in 1D \cite{Kozyreff-2006,Chapman-2009,Dean-2011}. Indeed, supposing that $X_0$ is above the real line and that $\Delta k<0$, let us examine the region of the complex plane given by $X=X_0-ir+\xi$, with $|\xi|\ll r$. With the aid of Stirling's approximation, $\Gamma(z+\alpha)\sim\sqrt{2\pi}z^{z+\alpha-1/2}e^{-z}$, (\ref{ansatz}) locally becomes
\begin{multline}
\frac{\eps^nb_{\ve q}\sqrt{2\pi} n^{\alpha-1/2}n^ne^{-n}e^{\ii\ve q\cdot\ve x}\ve u_T}{\left[-\Delta k(r+\ii\xi)\right]^n}\\
\sim
b_{\ve q}\sqrt{2\pi} n^{\alpha-1/2}\left(\frac{\eps n}{-\Delta k r}\right)^ne^{-n}e^{\ii\ve q\cdot\ve x-\ii n\xi/r}\ve u_T.
\end{multline}
This expression is smallest for $n=N\sim-\Delta k r/\eps$. Inserting this value and using the fact that $\xi=X-X_0+\ii r=\eps x-X_0+\ii r$, we obtain
\beq
b_{\ve q}\sqrt{2\pi} N^{\alpha-1/2}e^{\ii\ve q\cdot\ve x+\ii\Delta kx}\times e^{-i\Delta k X_0/\eps}\ve u_T.
\label{late}
\eeq
Since (\ref{expansion}) is truncated at $N^\text{th}$ order, we conclude that the remainder $R_N$ will contain terms proportional to (\ref{late}). This is fully analogous to the 1D case. As we have seen, $\ve q$ and $\Delta k$ above are such that $\exp\ii(\ve q\cdot\ve x+\Delta kx)$ is in fact one of the dominant Fourier modes of the hexagonal pattern. In the end, all the appropriate sets of $\ve q$, $\Delta k$, $X_0$ and its complex conjugate $\bar X_0$ give together rise to an extra, exponentially small contribution to the pattern given by (\ref{leading}). It is that contribution which allows the front to be stationary over a finite range of parameters \cite{Kozyreff-2006,Chapman-2009}. Hence, its size determines the width of the pinning range. With (\ref{late}), we have found contributions to the pinning range that are proportional to 
\beq
e^{-|\Delta k \Im(X_0)|/\eps},
\label{main}
\eeq
where $\Im(X_0)$ is the imaginary part of $X_0$. 

\begin{figure}
\includegraphics[width=7cm]{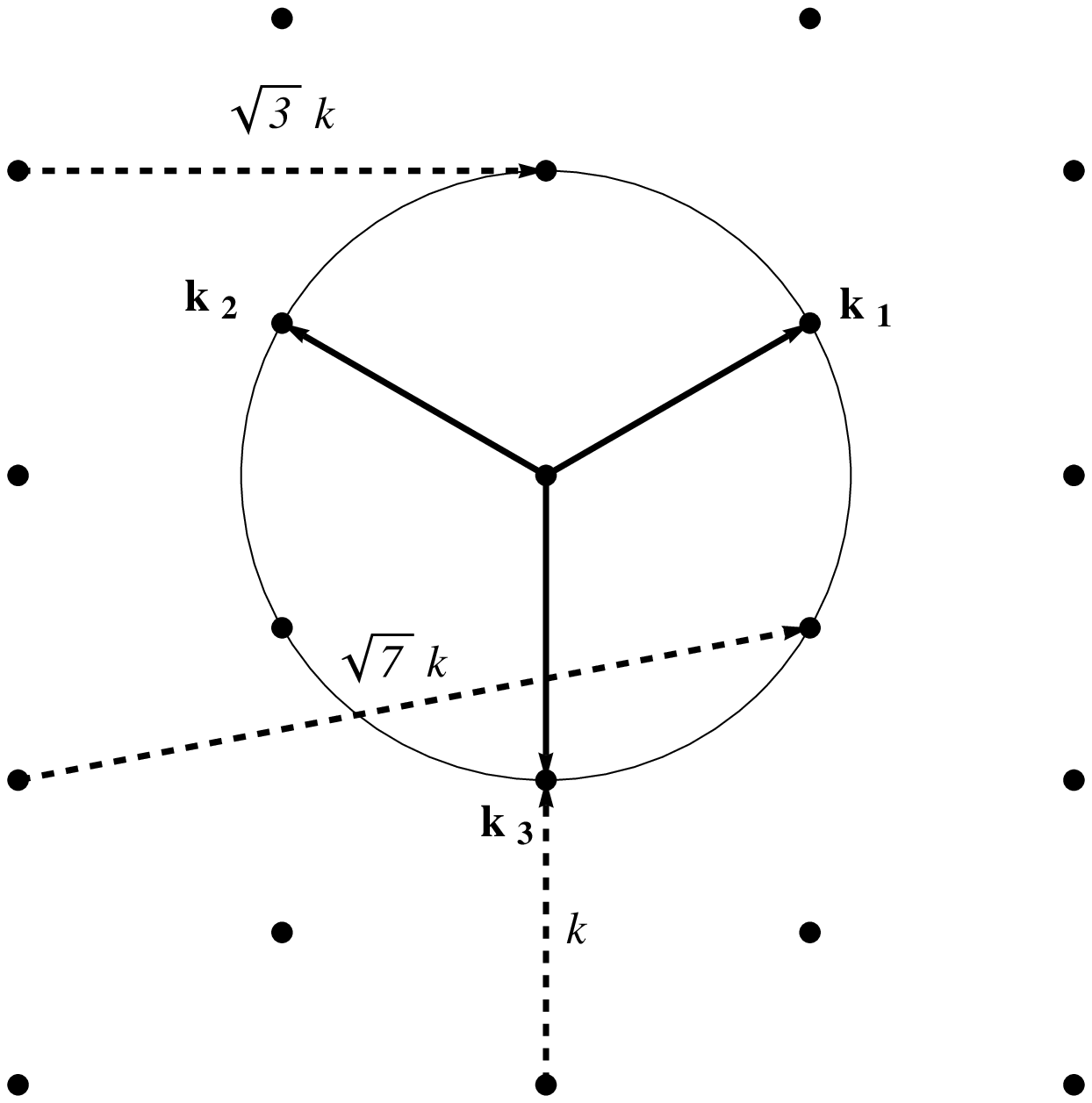}\\
\includegraphics[width=7cm]{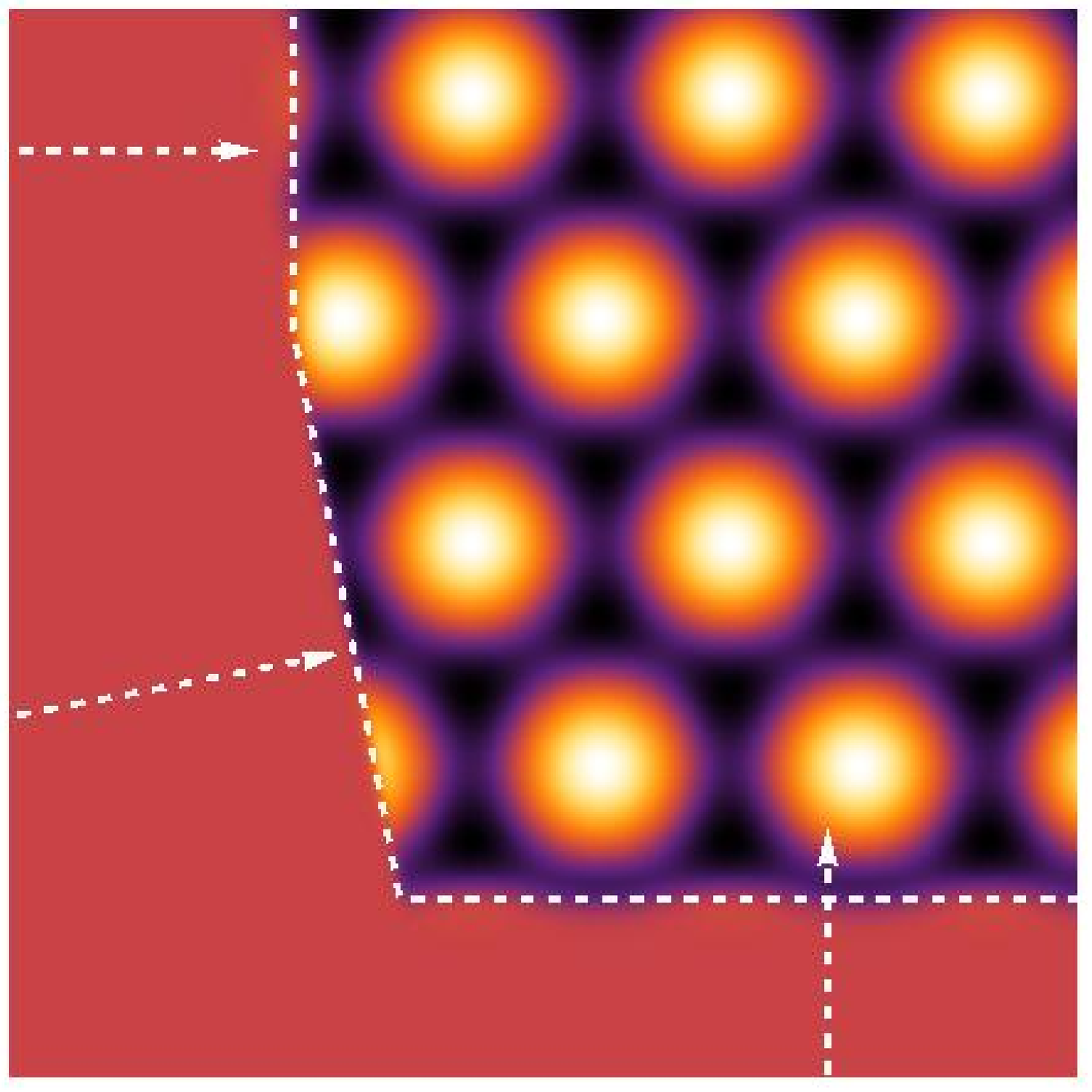}
\caption{(Color online) Top: the lattice in Fourier plane generated by the vectors $\ve k_i$ and three directions of translation symmetry (dotted arrows) along which fronts can lock. Bottom: The three corresponding fronts schematically depicted. The vertical arrow corresponds to the smallest $\Delta k$ and, hence, to the widest locking range.}
\label{fig:front}
\end{figure}

It remains to determine $\Im(X_0)$. We note that, in general, a front solution can be written $a(X)$, where $a$ is one of the real amplitudes $a_i$ and switches monotonically from $0$ to $\sqrt{2/5}$ as $X$ goes from $-\infty$ to $+\infty$. Hence we may in principle invert this relation and write $X=X(a)$. We are interested in the complex limit $X_0=\lim_{a\to\infty}X(a)$ or, more precisely, in its imaginary part. We know that, as $a\to\sqrt{2/5}$, we have $a\sim\sqrt{2/5}-p\exp(-\lambda X)$, where $\lambda$ is the smallest positive eigenvalue of (\ref{GL}) linearized around $a_i=\sqrt{2/5}$ and $p$ is some constant. Hence, locally, $X(a)\sim-\lambda^{-1} \ln(\sqrt{2/5}-a)$. By the same token, $a\sim q \exp(\mu X)$ near $a=0$ for some constants $q$ and $\mu$, implying the singularity $\mu^{-1}\ln(a)$ in $X(a)$. In the absence of other singularities in the function $X(a)$, we have
\beq
X(a)= \mu^{-1}\ln(a)-\lambda^{-1} \ln(\sqrt{2/5}-a)+\sum_{n\geq0} b_n a^n.
\eeq
It is easy to show, by linearization of (\ref{GL}), that $\mu$ and $\lambda$ are real. Hence, since $X$ is real on the real segment $0<a<\sqrt{2/5}$, the coefficients $b_n$ above are real too. If $a$ goes from 0 to $\infty$ through real values, it must go around the singularity at  $\sqrt{2/5}$. As it does so, $X(a)$ picks up an imaginary part given by $\pm(2l+1)\pi/\lambda$, where $l$ is the number of complete revolutions around the singularity, and where the + and  - signs correspond to clockwise and anti-clockwise rotation, respectively. Next, on the real interval $\sqrt{2/5}<\Re(a)<\infty$, only the real part of $X$ further changes, decreasing from infinity to a finite value. Hence, $\lim_{a\to\infty}\Im(X(a))=\Im(X_0)=\pm(2l+1)\pi/\lambda$. Evaluating (\ref{main}) with $l=0$, we thus obtain
\beq
\delta\nu(\eps)\propto e^{-\pi|\Delta k|/(\lambda \eps)}.
\label{main2}
\eeq
Finally, linearizing (\ref{GL}) about $a=\sqrt{2/5}$ for all admissible values $n_i$, we find that $0.669<\lambda< 0.675$, \textit{i.e.} that $\lambda^{-1}\approx1.5$. Hence, the pinning range approximately scales as
\beq
\delta\nu(\eps)\propto e^{-1.5\pi|\Delta k|/\eps},
\label{main3}
\eeq
where the dependence on front orientation only appears through $\Delta k$. Note that we use a proportionality sign, not `$\sim$', as we omit an algebraic factor involving some power of $\eps$. Expression (\ref{main3}) is the main result of this Letter. It shows that fronts for which $|\Delta k|$ is smallest have the largest pinning range. Considering Fig.~\ref{fig:front}, it appears that the smallest possible value of $|\Delta k|$ is $k$ and is found when the $x$ axis is parallel to one of the $\ve k_i$. In Miller's notation, this is a ``[10]" front \cite{Lloyd-2008} . The next smallest $|\Delta k|$ is $\sqrt3k$, and is obtained when the $x$ axis is perpendicular to one of the $\ve k_i$ (a ``[11]" front.) The third smallest is value is $\sqrt7k$, corresponding to the $x$ axis parallel to $\ve k_i-2\ve k_j$, $i\neq j$; this is  a ``[12]" front. These fronts are schematically depicted in Fig.~\ref{fig:front}.

To test our theory, we solved (\ref{SH}) and (\ref{NLO}) numerically with localized initial conditions and determined the parameter ranges in which localized patterns were stable. The stability region so computed is expected to accurately coincide with the pinning range if the localized patterns are more than 5 peaks wide \cite{Gomila-2007}. The size of the domain was taken to be large in the $x$ direction and exactly one period wide in the $y$-direction. By enforcing periodic boundary conditions and adjusting the domain size in the $y$-direction, we controlled the orientation of the hexagons with respect to the $x$ axis \cite{Avitabile-2010}. We investigated the parameter space suggested by the weakly nonlinear analysis and found confirmation that, as $\eps\to0$, the largest pinning range is found for fronts with $\Delta k=k$, followed by those with $\Delta k=\sqrt3k$ and then by the fronts with $\Delta k=\sqrt7k$, see  Figs.~\ref{PRLnumerics1} and \ref{PRLnumerics2}. 

\begin{figure}
\includegraphics[width=8cm]{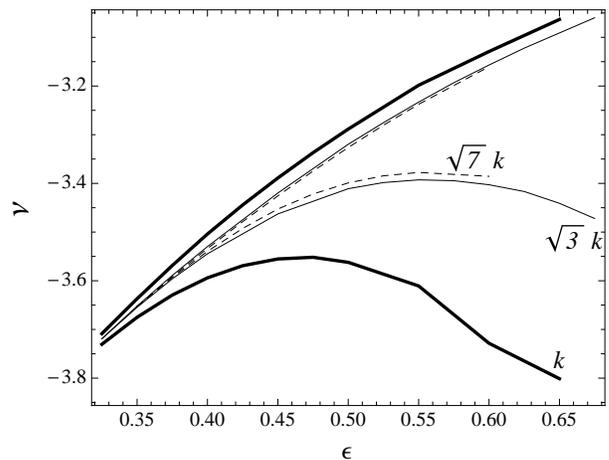}
\caption{Pinning region for Eq.~(\ref{SH}) with $r=-\eps^2$ and $s=\sqrt{3/4}\eps\nu$. Thick line: ``[10]" front, with $\Delta k =k$. Thin line: ``[11]" front, with  $\Delta k =\sqrt3k$. Dashed line: ``[12]" front, with  $\Delta k =\sqrt7k$. }
\label{PRLnumerics1}
\end{figure}
\begin{figure}
\includegraphics[width=8cm]{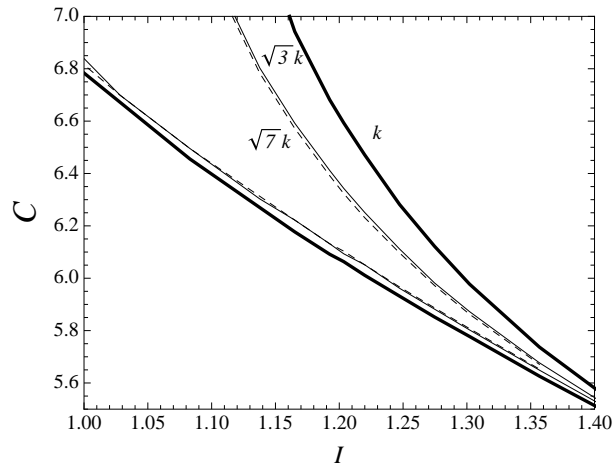}
\caption{Pinning region for Eq.~(\ref{NLO}). Thick line: ``[10]" front, with $\Delta k =k$. Thin line: ``[11]" front, with  $\Delta k =\sqrt3k$. Dashed line: ``[12]" front, with  $\Delta k =\sqrt7k$.}
\label{PRLnumerics2}
\end{figure}
Expressions (\ref{main}), (\ref{main2}) and (\ref{main3}) can be simply interpreted: The relevant length scale of the patterns in the $x$ direction is given by $4\pi/\Delta k$, while the length scale associated to the front is given by $1/\eps \lambda$. The true small parameter of the problem is given by the ratio of the two. The strength of the pinning between the front and the underlying pattern is exponentially small in that parameter. This geometrical argument and the reasoning that lead to (\ref{main3}) do not depend on special model features. The main assumptions of the theory are (\textit{i}) separation of spatial scales, (\textit{ii}) weak nonlinearity. The latter ensures that the solution is composed of well defined and regularly spaced peaks in the Fourier plane. In principle, the same argument could be carried for square patterns and 3D patterns.  Note that the model need not be variational for the present results to hold. Indeed, Eq.~(\ref{NLO}) does not derive from a potential.

 Systematic calculations of curves like those shown in Figs.~\ref{PRLnumerics1} and \ref{PRLnumerics2} are rare. Previous calculations have been made for the ``[10]" and ``[11]" fronts in the Swift-Hohenberg model \cite{Lloyd-2008}; they agree with the present results. More recent calculations of hexagonal patches in an urban crime model \cite{Lloyd-2013} also agree with the above conclusions, even though the limit $(\eps,\nu)=(0,-\sqrt{45/2})$ cannot be reached with the available control parameters of that model.

When the front is normal to one of the $\ve k_i$ or, equivalently, when it is parallel to one of the sides of the elementary hexagon, its pinning force is the strongest. Consequently, slightly outside the pinning region, this type of front is  the more persistent and more likely to be observed. Growth or decay in all other directions is exponentially faster. Hence, evolving localized patterns just outside the pinning range tend to have polygonal shapes with their sides (the fronts) parallel to the sides of the elementary hexagonal cell. This sheds new light on previous numerical simulations on chemical patterns \cite{Jensen-1994}, optical patterns \cite{Harkness-2002}, and the Swift-Hohenberg equation \cite{Boyer-2003}, as well as gazeous CO$_2$  experiments \cite{Bodenschatz-1991}.

\acknowledgments
G.K. is a Research Associate of the Fonds de la Recherche Scientifique - FNRS (Belgium.) G.K. thanks Pascal Kockaert for useful comments. This research was supported in part by the Interuniversity Attraction Poles program of the Belgian Science Policy Office under Grant No. IAP P7-35.

%

\end{document}